\documentclass[]{spie}  

\usepackage{amsmath,amsfonts,amssymb}
\usepackage{graphicx}
\usepackage[colorlinks=true, allcolors=blue]{hyperref}

\title{GroundBIRD: First On-Sky Responsivity Calibration at the Teide Observatory} 

\author[a,b]{Alessandro Fasano}
\author[c]{Yonggil Jo}
\author[a,b]{Ricardo Tanausú Génova-Santos}
\author[d]{Makoto Hattori}
\author[e,f]{Shunsuke Honda}
\author[g]{Kenichi Karatsu}
\author[h,i]{Chiko Otani}
\author[j]{Michael Peel}
\author[a,b]{José Alberto Rubiño-Martín}
\author[k]{Yoshinori Sueno}
\author[l]{Junya Suzuki}
\author[k,l,m]{Osamu Tajima}
\author[d]{Tomonaga Tanaka}
\author[d]{Miku Tsujii}
\author[c]{Eunil Won}

\affil[a]{Instituto de Astrofísica de Canarias, E-38200 La Laguna, Tenerife, Spain}
\affil[b]{Departamento de Astrofísica, Universidad de La Laguna (ULL), E-38206 La Laguna, Tenerife, Spain}
\affil[c]{Department of Physics, Korea University, Seoul, South Korea}
\affil[d]{Astronomical Institute, Tohoku University, Sendai 980-8578, Japan}
\affil[e]{Division of Physics, Faculty of Pure and Applied Sciences, University of Tsukuba, Tsukuba, Ibaraki 305-8571, Japan}
\affil[f]{Tomonaga Center for the History of the Universe (TCHoU), Faculty of Pure and Applied Sciences, University of Tsukuba, Tsukuba, Ibaraki 305-8571, Japan}
\affil[g]{Space Research Organization Netherlands (SRON), Niels Bohrweg 4, Leiden, 2333 CA, The Netherlands}
\affil[h]{Tohoku University, 2-1-1 Katahira, Aoba-ku, Sendai, Miyagi 980-8577, Japan}
\affil[i]{RIKEN, 519-1399 Aramaki-Aoba, Aoba-ku, Sendai, Miyagi 980-0845, Japan}
\affil[j]{Imperial College London, South Kensington Campus, London SW7 2AZ, UK}
\affil[k]{Kavli IPMU (WPI), UTIAS, The University of Tokyo, Kashiwa, Chiba 277-8583, Japan}
\affil[l]{Department of Physics, Faculty of Science, Kyoto University, Kyoto 606-8502, Japan}
\affil[m]{High Energy Accelerator Research Organization (KEK), Tsukuba 305-0801, Japan}

\authorinfo{Further author information: (Send correspondence to Alessandro Fasano)\\Alessandro Fasano: E-mail: alessandro.fasano@iac.es}

\begin{document} 
\maketitle

\begin{abstract}
GroundBIRD is a cosmic microwave background (CMB) experiment located at the Teide Observatory (altitude $\sim2\,400$\,m, Spain) that is designed to measure large-angular-scale intensity and polarization anisotropies ($\ell \gtrsim 6$ up to $\ell \sim 300$) with the goal to constrain the optical depth to reionization, $\tau$. The instrument employs a rapidly rotating telescope with the elevation fixed at $70$\,deg and is equipped with 161 lenslet-coupled kinetic inductance detectors (KIDs): 138 KIDs at 145\,GHz dedicated to CMB observations and 23 KIDs at 220\,GHz for thermal dust characterization. All detectors are operated at a temperature of $\sim280$\,mK. This scan strategy provides daily coverage of approximately 40\% of the Northern Hemisphere sky.

We present the first on-sky responsivity calibration model for GroundBIRD, derived from repeated observations of Jupiter using the most stable subset of twelve KIDs from GroundBIRD array 6 (GB06). The analysis focuses on developing and validating a calibration methodology on this restricted subset of detectors and yields the first empirical characterization of detector responsivity as a function of precipitable water vapor (PWV). We find that the detector responsivity decreases by approximately 30\% across the sampled PWV range and is well described by a linear PWV-dependent model under typical observing conditions. A comparison with observations of the Moon reveals the existence of distinct detector operating regimes associated with different optical loading conditions. This analysis demonstrates that the detector responsivity is governed by the total optical load, rather than exclusively by the atmospheric contribution to the optical load.

The Jupiter-based calibration achieves a precision better than 20\% under typical observing conditions and exhibits a relative stability better than 20\% over a one-month timescale. These results underscore the critical importance of accounting for atmospheric effects in responsivity calibration and provide the foundation for extending the calibration methodology to the full GroundBIRD focal plane.

\end{abstract}


\section{INTRODUCTION}
\label{sec:intro}  

Measurements of large-scale anisotropies in the cosmic microwave background (CMB) polarization provide a direct probe of the optical depth to reionization, $\tau$.\cite{2020A&A...641A...6P,2025ApJ...986..111L} GroundBIRD\cite{2012SPIE.8452E..1MT} is designed to address this goal by performing wide-area measurements of CMB intensity and polarization at degree angular scales ($\ell \gtrsim 6$ up to $\ell \sim 300$) from the Teide Observatory (altitude $\sim2\,400$\,m, 28$^\circ$18'01.8''N 16$^\circ$30'37''W, Spain),\cite{2021ApJ...915...88L} using a rapidly rotating telescope\cite{2018JLTP..193.1066N} and a focal plane of kinetic inductance detectors (KIDs),\cite{2003Natur.425..817D} 138 distributed among six arrays at 145\,GHz and 23 in one array at 220\,GHz, all operating at $\sim280$\,mK.\cite{2024SPIE13102E..05T}

Establishing an on-sky responsivity model is essential for astronomical observations with KID-based instruments. The model relates the detector readout to the incident sky signal while accounting for the telescope and optical chain. Because the detector operating point under observing conditions typically differs from that achievable in the laboratory, the relevant calibration parameters must be determined directly from astronomical data.

In this work, we present the first on-sky responsivity model developed for GroundBIRD. Using Jupiter observations acquired under varying atmospheric conditions, we derive the first empirical precipitable water vapor (PWV) dependent responsivity calibration for the instrument. By comparing calibrations derived from Jupiter and Moon observations, we additionally identify two distinct detector operating regimes associated with different optical-loading conditions. These results establish the foundation for future full-focal-plane calibration of GroundBIRD.

Section~\ref{sec:responsivity} introduces the KIDs response framework.
We address the Jupiter-based responsivity calibration in Sect.~\ref{sec:jupiter}, including the scan selection and methodology (Sect.~\ref{sec:method}), the relative-response characterization (Sect.~\ref{sec:relative_response}), the empirical PWV-dependent calibration model (Sect.~\ref{sec:responsivity_vs_pwv}), and the beam-stability analysis (Sect.~\ref{sec:gb06_stability}). 
In Sect.~\ref{sec:moon}, we present Moon- and Jupiter-derived responsivities and investigate the dependence of the detector response on lunar illumination (Sect.~\ref{subsec:moon_vs_jupiter}) and on the total optical load (Sect.~\ref{sec:total_optical_load}). We then (Sect.~\ref{sec:jupiter_as_calibrator}) discuss the validity of Jupiter as a calibrator of the atmosphere-dominated operating regime relevant for nominal GroundBIRD observations.
Finally, Sect.~\ref{sec:conclusions} summarizes the main results and outlines future developments.

\section{KIDS RESPONSE FRAMEWORK}
\label{sec:responsivity}

Kinetic inductance detectors are superconducting resonators whose resonance properties vary in response to absorbed radiation through the generation of quasiparticles (see Ref.~\citenum{kids} for a detailed review). Due to their intrinsic frequency-domain multiplexing capability, high sensitivity, and scalability to large-format arrays, KIDs have evolved into a mature and widely adopted technology for observations in the millimeter (mm) and sub-mm wavelength regimes of astronomy.\cite{2008JLTP..151..530D}

Although the KID physics is intrinsically linear over a broad dynamic range,\cite{2014JLTP..176..787M} the effective responsivity inferred from phase readout is dependent on the detector’s operating point. Changes in optical and thermal loading alter both the resonance frequency and the quality factor, thereby modifying the mapping between the absorbed optical power and the measured phase signal.\cite{Gao,kids,2015ApPhL.106g3505H}

Before each observation, the resonance frequency of every KID is determined through a tuning procedure that employs a frequency sweep in the in-phase and quadrature (I,Q) complex plane (see, e.g., Ref.~\citenum{2022JInst..17P8037B} for a detailed description of a state-of-the-art KID tuning methodology). 
During standard survey operations, GroundBIRD executes this tuning procedure at 10\,minute intervals. In contrast, dedicated calibration-source observations generally adopt a 1\,hour tuning cadence to maximize the on-source integration time. The KIDs are then read out at a fixed frequency by recording the complex (I,Q) signal, which is subsequently converted into amplitude and phase time streams.

Temporal and spatial variations in atmospheric emission induce shifts in the KID resonance frequency and changes in the quality factor. When the resonance frequency drifts sufficiently far from the fixed readout tone, the detector operates in an off-resonance regime, in which the phase response exhibits an effectively nonlinear behavior.\cite{fasano_aa} This apparent nonlinearity does not originate from any intrinsic nonlinear properties of the KID itself, but rather from the specifics of the readout geometry and the evolution of the detector’s operating point.

Several techniques, including 2-point and 3-point modulation, have been developed to reconstruct the resonance-frequency evolution in real time and thereby mitigate responsivity variations associated with changing operating conditions.\cite{Calvo2013,fasano_aa,2022JInst..17P8037B} GroundBIRD currently relies on fixed-frequency phase readout, making an empirical on-sky responsivity calibration necessary.

GroundBIRD conducts observations using a continuous azimuthal spin scan at a fixed elevation angle of $70$\,deg. This scan strategy enables daily coverage of $\sim40\%$ of the Northern Hemisphere sky while preserving a constant air mass throughout the observations. Scientific data acquisition is structured as repeated cycles, each consisting of a KID tuning procedure followed by fixed-bias frequency-time-domain measurements. Under nominal observing conditions, the tuning cycle is repeated every 10\,minutes to compensate for typical variations in the atmospheric background at the Teide Observatory.\cite{11429154}
For observations targeting calibration sources, the tuning interval is extended to 1\,hour to mitigate data loss, at the cost of a reduced ability to track variations in the KID operating point between retuning operations.
The impact of this effect on the responsivity calibration is discussed in Sect.~\ref{sec:jupiter_as_calibrator}.

\section{RESPONSIVITY CALIBRATION}
\label{sec:jupiter}

As a bright, effectively point-like source with respect to the GroundBIRD beam (full width at half maximum; FWHM$\sim30$\,arcmin at 145\,GHz),\cite{2026ITAS...36S2265T} Jupiter provides a stable reference for characterizing the detector response. 
The calibration is developed from the measured Jupiter amplitudes after correcting for atmospheric absorption and the time-dependent beam dilution associated with Jupiter's varying angular diameter.
In the following sections, we describe the scan selection, define the detector and array responsivities, and derive an empirical model of their dependence on PWV. The validity of Jupiter as a calibrator is further assessed in Sect.~\ref{sec:jupiter_as_calibrator}.

In this study, atmospheric conditions are monitored using the PWV measured by a Furuno Electric Co., Ltd.\footnote{\url{https://www.furuno.com/}} microwave radiometer installed $\sim40$\,m from the telescope. The instrument observes the 22\,GHz water-vapor line at zenith with a 16\,deg FWHM beam and provides PWV estimates every $\sim17$\,s.

\subsection{Scan Selection and Methodology}
\label{sec:method}

In this work, the responsivity calibration is derived from eight\footnote{1, 6, 7, 20, 21, 22, 23, and 27 February 2024.} Jupiter observations conducted throughout February 2024.
The analysis is restricted to KIDs exhibiting a clearly identifiable Jupiter signal in multiple observations over the course of the campaign. 
To obtain a statistically robust characterization of both responsivity and temporal stability, each KID is required to exhibit a detectable Jupiter signal in at least three independent scans.

Among the six GroundBIRD arrays operating at 145\,GHz, GroundBIRD array 6 (GB06) was selected for detailed analysis because it provided the most reliable detector characterization throughout the entire campaign. In particular, GB06 exhibited the cleanest resonance sweeps, enabling stable and reproducible resonance fits, and yielded the largest number of KIDs with robust Jupiter detections across multiple observations. The KIDs in GB06 are clustered around an elevation of $65.0\pm1.1$\,deg, where the quoted dispersion corresponds to the spread in the reconstructed beam centers of the individual detectors within the array.
The remaining arrays are excluded from the primary analysis because a substantial fraction of their detectors were affected by known issues identified during the commissioning period. These issues include unstable or poorly fitted resonance sweeps, increased susceptibility to microphonic contamination, intermittent readout instabilities, and reduced signal-to-noise ratio (SNR) in the Jupiter maps.

For each Jupiter observation (with a typical duration of $\sim1$\,hour), the corresponding map was modeled using a circular two-dimensional Gaussian profile plus a constant baseline term. 
The best-fit peak amplitude for each KID, $A_{\rm k}$, was adopted as a measure of the detector response, while the best-fit Gaussian width was used to characterize the beam FWHM.
To assess the quality of the reconstructed maps, we estimated the SNR via aperture photometry. 
The signal was defined as the background-subtracted integrated flux within a circular aperture centered on the best-fit source position and with a radius equal to half of the fitted FWHM. The background level and root-mean-square fluctuation were estimated from a concentric, source-free annulus centered on the same position, with an inner radius of twice the fitted FWHM and an outer radius chosen such that the annulus enclosed the same area as the source aperture. 

\begin{figure}[ht]
    \centering
    \includegraphics[width=0.95\linewidth]{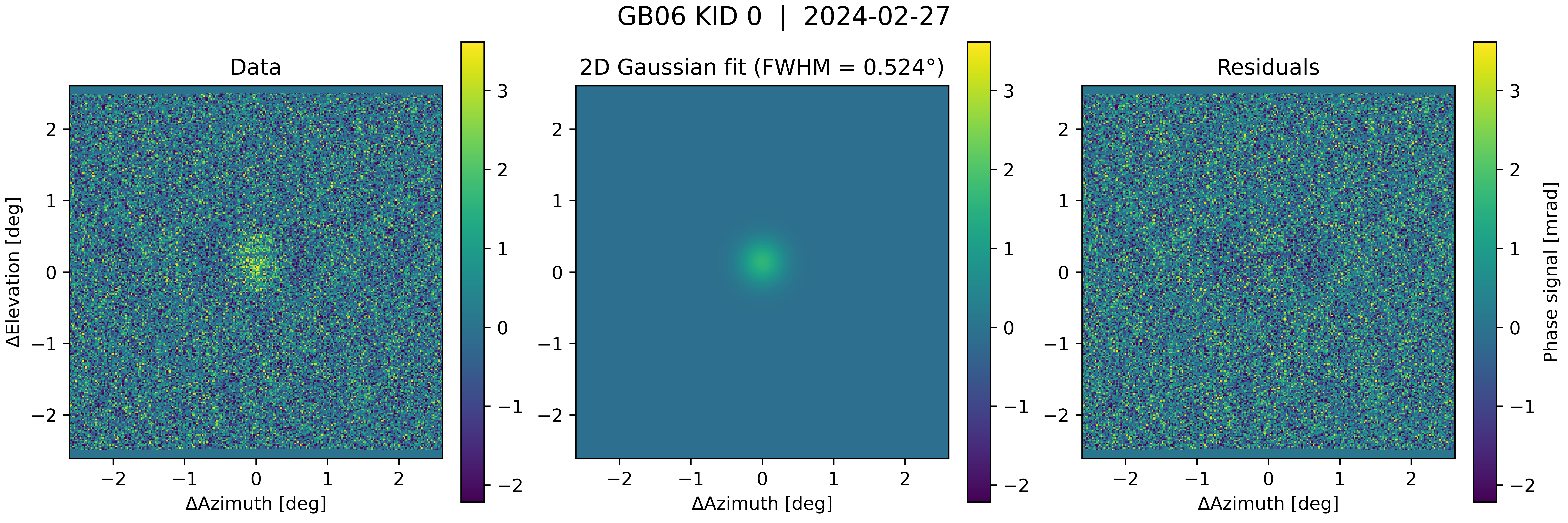}
    \caption{Example of a Jupiter map derived from observations with a GB06 detector (GB06 KID 0) during a one-hour fixed-elevation scan conducted on 27 February 2024. The left panel displays the reconstructed signal map, the central panel shows the best-fit circular two-dimensional Gaussian model, and the right panel presents the residuals obtained after subtracting the model from the data. The maps are represented in differential elevation–azimuth coordinates, centered on the predicted position of Jupiter. The color scale indicates the uncalibrated detector phase response in mrad. In this observation, Jupiter is detected with a signal-to-noise ratio $\simeq17.8$.}
    \label{fig:example_jupiter_fitting}
\end{figure}

The SNR was then defined as the ratio of the integrated signal to the corresponding aperture noise. The resulting sample exhibits a mean SNR of $15.5\pm3.9$, where the quoted uncertainty corresponds to the standard deviation over all successfully reconstructed Jupiter observations.
An illustrative example of a reconstructed Jupiter map, together with the corresponding best-fit model and residuals, is shown in Fig.~\ref{fig:example_jupiter_fitting}.

The Jupiter brightness temperature was set to
$T_{\rm B}=170.7$\,K, following the results reported in Ref.~\citenum{2017A&A...607A.122P}. 
This brightness temperature was converted to an apparent antenna temperature by accounting for Jupiter's time-dependent angular diameter and beam dilution, using the fitted beam FWHM derived from each KID map. The resulting quantity, $T_{\rm A,k}(t)$, therefore represents the expected antenna temperature observed by detector k at epoch $t$.
The signal was corrected for atmospheric absorption using the measured PWV and the \texttt{am} atmospheric transmission model.\cite{paine_scott_2022_6774376} 
The model was configured with the GroundBIRD site parameters, pointing model per KID, and the instrumental bandpass derived from the manufacturer-provided transmission curves of the optical filter chain.
The KID responsivity was then defined as
\begin{equation}
\mathfrak{R}_{\rm k}(t) =
\frac{A_{\rm k}^{\mathrm{corr}}(t)}
{T_{ \rm A,k}(t)} \, ,
\label{eq:calibration}
\end{equation}
\noindent where $A_{\rm k}^{\mathrm{corr}}(t)$ is the opacity-corrected Jupiter signal amplitude in randians and $T_{ \rm A,k}(t)$ is the corresponding model antenna temperature.
Astrophysical signals are subsequently calibrated into antenna-temperature units using the measured KID responsivity, in other words, by inverting Eq.~\ref{eq:calibration}.
The $\mathfrak{R}_{\rm k}(t)$ median value for each detector across the campaign, $
\langle\mathfrak{R}_{\rm k}\rangle$, is reported in Appendix~\ref{app:relative_response}, in Tab.~\ref{tab:flatfield}.

\subsection{KIDs Relative Response and Array Responsivity}
\label{sec:relative_response}

The responsivities derived from Jupiter observations exhibit significant KID-to-KID variations, as expected from the underlying detector physics. Differences in resonator characteristics, optical coupling, and readout conditions naturally lead to variations in the absolute responsivity of individual detectors. To isolate the temporal evolution of the array's global responsivity from these predominantly static detector-dependent effects, we define a relative response for each KID ($F_{\rm k}$), commonly referred to as a flat field.

\begin{figure}[ht]
    \centering
    \includegraphics[width=0.72\linewidth]{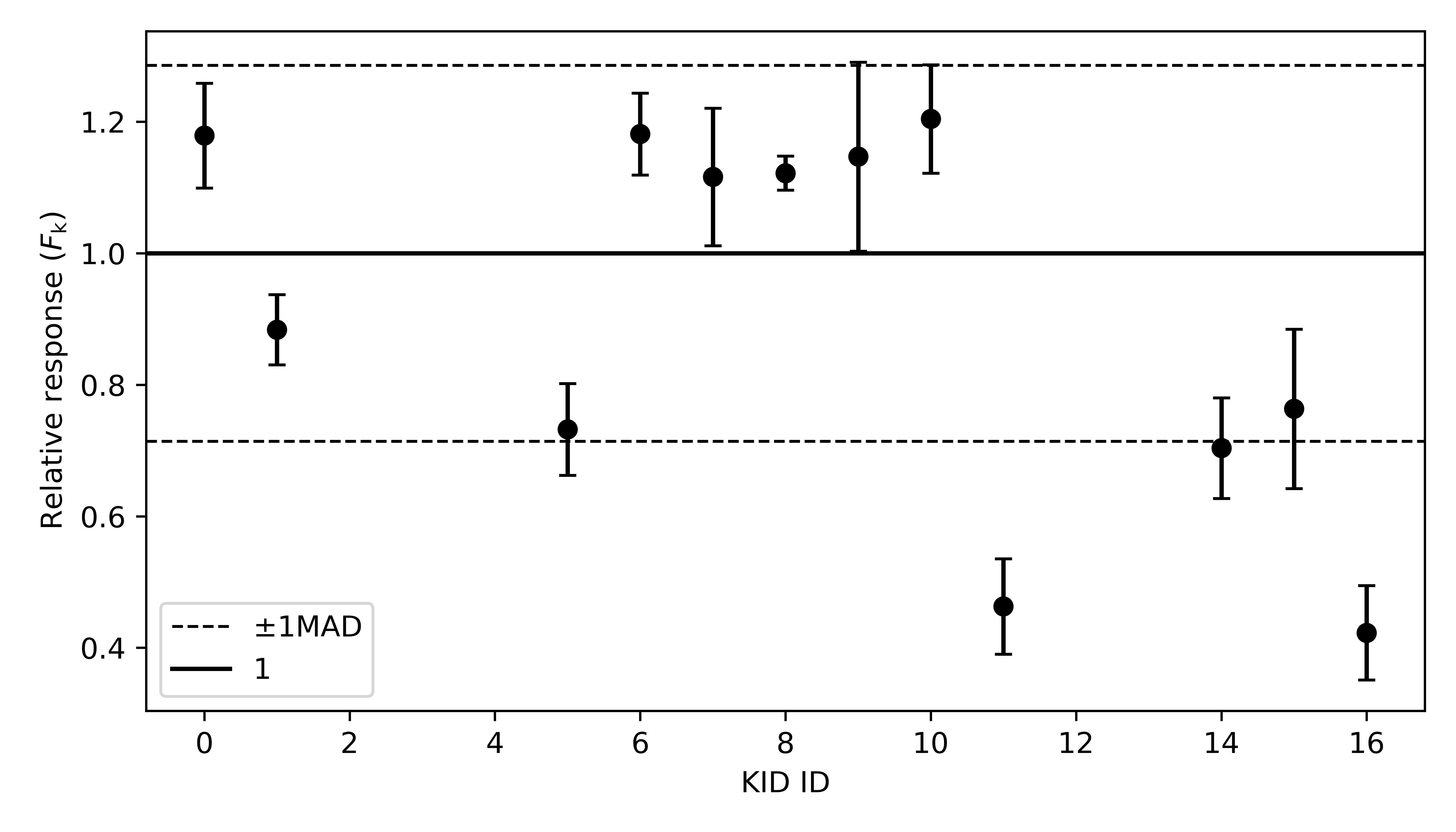}
    \caption{
    Relative response ($F_{\rm k}$) for the selected GB06 detectors (identified by their identification number, KID ID) derived from monthly Jupiter observations. Error bars represent the temporal scatter of the responsivity measurements. The horizontal line marks the normalized reference level ($F_{\rm k}=1$), while the dashed lines indicate the median absolute deviation (MAD) of the KID ensemble ($\pm0.29$). We report the results in Tab.~\ref{tab:flatfield}.
    }
    \label{fig:gb06_relative_response}
\end{figure}

Using the selected detector ensemble, a reference array responsivity is computed at each observing epoch as the median responsivity across all available detectors. The relative response for each KID is then estimated as the median ratio of its responsivity to the reference array responsivity over the full observing period. Finally, the coefficients are normalized so that their median is unity.
This procedure removes static KID-to-KID gain differences and places all detectors on a common responsivity scale. The resulting relative response coefficients are shown in Fig.~\ref{fig:gb06_relative_response}.

Each selected KID exhibits month-scale responsivity stability better than 20\%. 
The residual scatter among detectors ($\pm$0.29) reflects intrinsic KID-to-KID variations in gain and optical coupling. 
Appendix~\ref{app:relative_response} reports the values of the relative response per detector.

To preserve KID-level information while extracting the ensemble's common calibration, the individual responsivities are combined after normalization using their relative response coefficients. The array responsivity is, thus, defined as

\begin{equation}
\mathfrak{R}(t)=
\mathrm{median}_{k\in\mathcal{S}}
\left[
\frac{\mathfrak{R}_k(t)}{F_k} 
\right] \, ,
\end{equation}
\noindent where $S$ denotes the ensemble of the selected KIDs.
The median provides a robust estimate of the global detector response, minimizing the impact of residual outliers and occasional KID instabilities. The corresponding uncertainty is estimated from the KID-to-KID scatter using the scaled median absolute deviation.

\subsection{Empirical PWV Dependence}
\label{sec:responsivity_vs_pwv}

Having established a robust estimate of the array responsivity, we now investigate its dependence on atmospheric conditions. At mm wavelengths, the dominant contribution to the variable optical background is atmospheric emission, primarily driven by fluctuations in PWV. Variations in PWV modify both the atmospheric transmission and the total optical loading that is incident on the detectors.\cite{2022PhRvD.105d2004M,2025PhRvD.111h2001M}

\begin{figure}[ht]
    \centering
    \includegraphics[width=0.72\linewidth]{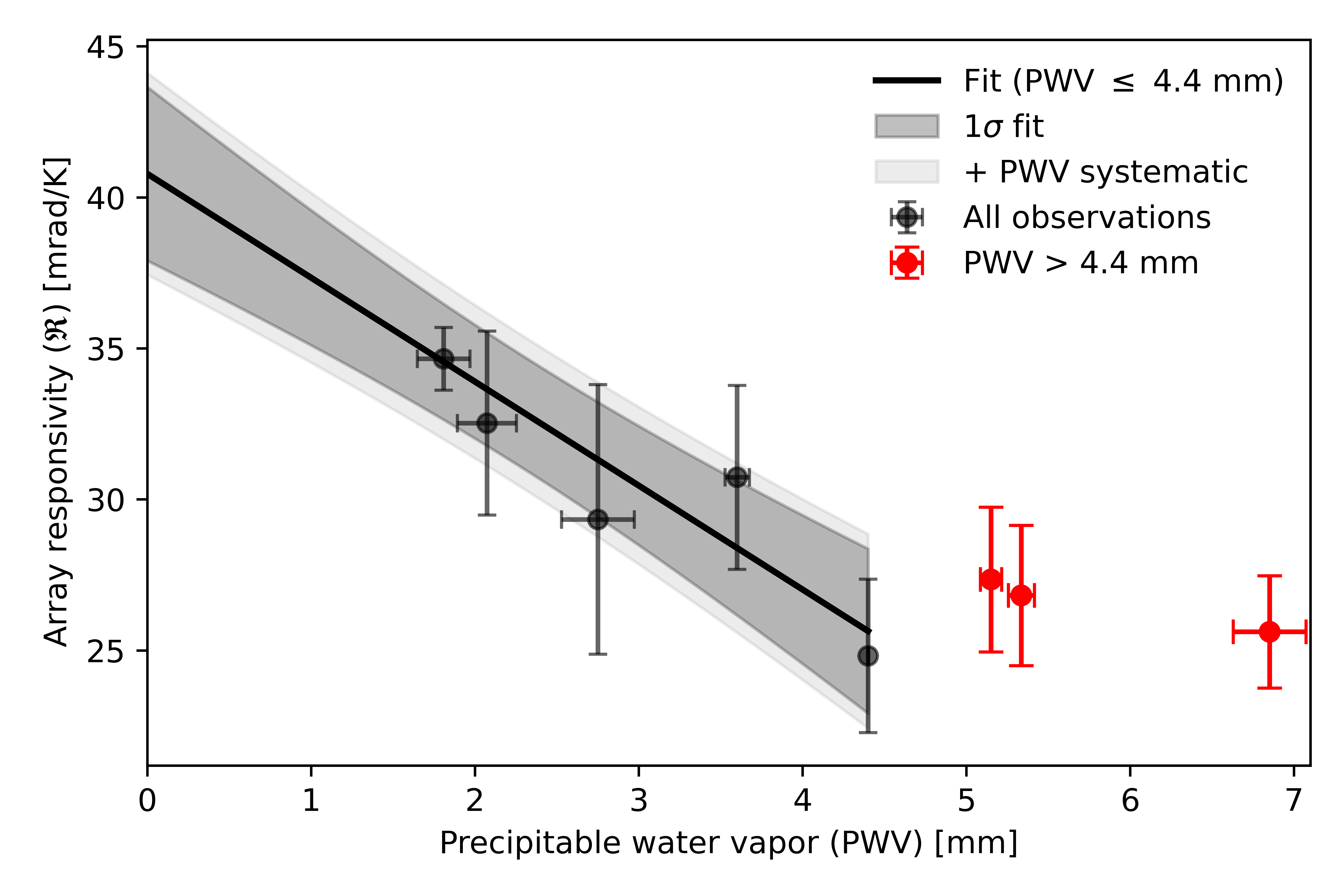}
    \caption{
    Array responsivity as a function of PWV. Error bars show the detector-to-detector scatter estimated from the scaled median absolute deviation. The black points show all observations, while the red points highlight measurements obtained at $\mathrm{PWV}>4.4\,\mathrm{mm}$. The black line shows the linear fit performed using only the low-PWV regime ($\rm{PWV}\leq4.4\,\mathrm{mm}$). The best-fit relation is $\mathfrak{R}(\rm{PWV})=40.8\pm2.4\,\mathrm{mrad/K} + (-3.4\pm1.0\,\mathrm{mrad/K/mm})\,\rm{PWV} $. The dark shaded region represents the statistical $1\sigma$ fit uncertainty, while the lighter region additionally includes the systematic uncertainty associated with PWV fluctuations during a typical observing scan of one hour (0.7\,mm).\cite{11429154}
    }
    \label{fig:gb06_resp_pwv}
\end{figure}

An increase in atmospheric background shifts the detector operating point and changes the local slope of the resonance. As a result, the same Jupiter antenna temperature can yield different phase responses under varying atmospheric conditions.
By comparing the responsivity derived from each Jupiter observation with the contemporaneous PWV measurement, we can empirically relate the impact of atmospheric conditions on the array responsivity. The result is shown in Fig.~\ref{fig:gb06_resp_pwv}.

Over the PWV range sampled by the observations ($\sim$1.8--4.4\,mm), the array responsivity decreases by $\sim30\%$.
The linear fit was performed using the five observations with PWV $\leq 4.4\,\mathrm{mm}$. The resulting model provides an adequate description of the data, with a reduced chi-squared of $\chi^2_\nu = 0.35$. The responsivity is anti-correlated with PWV, with a Pearson correlation coefficient of $r=-0.90$ and a Spearman rank correlation coefficient of $\rho=-0.90$. The fitted slope is statistically different from zero, with a two-sided p-value of 0.036.
Despite the limited size of the statistical sample, the data show a trend toward lower responsivities at higher atmospheric loading, consistent with the expected behavior of KIDs.
The linear fit yields an intercept of $\mathfrak{R}_0 = 40.8 \pm 2.4\,\mathrm{mrad/K}$ and a slope of $\alpha = -3.4 \pm 1.0\,\mathrm{mrad/K/mm}$.
Under the observing conditions encountered during the GroundBIRD campaign at Teide Observatory between July 2023 and August 2024 (PWV $\sim 4.3\,\mathrm{mm}$),\cite{11429154} the PWV-based preliminary responsivity calibration predicts the responsivity with a statistical uncertainty better than $20\%$.

As shown in Fig.~\ref{fig:gb06_resp_pwv}, the responsivity varies approximately linearly with PWV for PWV$\leq$4.4\,mm. Above this value, the responsivity shows systematic deviations from the low-PWV trend, consistent with the KIDs entering a higher-loading regime in which changes in resonance properties become increasingly dominant and a simple linear description is no longer sufficient. The value PWV=4.4\,mm does not represent a precisely determined physical transition; rather, it corresponds to the highest PWV observation still consistent with the low-PWV linear relation. Since all measurements at larger PWV show systematic departures from this trend, we adopt PWV=4.4\,mm as a conservative upper limit for the validity of the linear calibration model and for routine GroundBIRD observations.

Although only twelve of the twenty-three KIDs in GB06 satisfy the adopted selection criteria, the array responsivity is determined from the median response of the full detector ensemble. It remains stable throughout the observing campaign. The observed calibration trends are significantly larger than the residual detector-to-detector scatter, demonstrating that the selected subset is sufficient to both establish and validate the calibration methodology.  
The statistical significance of the derived trend is constrained primarily by the limited number of available Jupiter observations, rather than by the intrinsic KID-to-KID variability.  
The same analysis framework is currently being applied to additional Jupiter observations acquired whenever scheduling and atmospheric conditions permit. These data will be used to extend the calibration to the full focal plane and to improve the statistical characterization of the KID responsivities over a broader range of atmospheric conditions. 

The calibration framework developed in this work can be further validated through observations of additional planetary targets. In particular, Venus represents an attractive validation source, as its inferred brightness temperature can be directly compared with measurements obtained by other instruments operating at similar frequencies, including \textit{KISS} at the Teide Observatory\cite{fasano_aa} and \textit{CLASS} at the Chajnantor Plateau.\cite{2023PSJ.....4..154D} Using the measured GroundBIRD beam profile in combination with planetary ephemerides, we estimate that Venus should produce a beam-diluted signal with an amplitude of $\sim40\%$ of that measured for Jupiter.

Observations of Saturn would provide a complementary consistency test. The same analysis predicts a beam-diluted signal with an amplitude of $\sim20\%$ of the Jupiter signal. Although significantly fainter than Jupiter, Saturn offers the advantage that its emission can be directly compared with the planetary brightness-temperature measurements from the \textit{Planck} mission reported in Ref.~\citenum{2017A&A...607A.122P}.

Furthermore, dedicated skydip observations\cite{1981ApJ...250..341K} could be used to directly characterize the dependence of detector responsivity on atmospheric loading. Such measurements would provide an independent validation of the PWV-dependent responsivity model and help disentangle the respective contributions of atmospheric and source loading to the detector operating point.

\subsection{Beam Stability}
\label{sec:gb06_stability}

\begin{figure}[ht]
    \centering
    \includegraphics[width=0.72\linewidth]{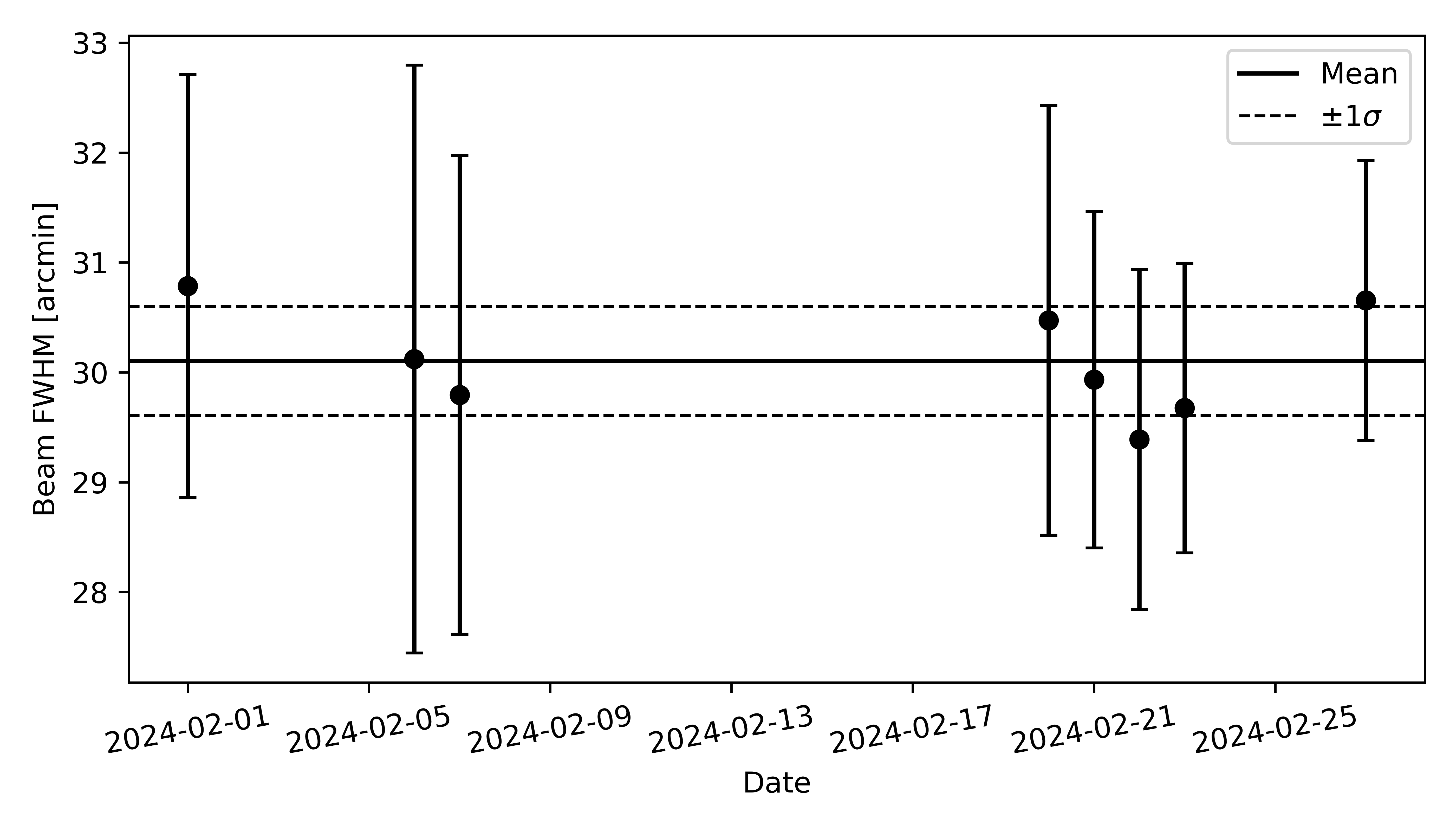}
    \caption{
Median beam full width at half maximum (beam FWHM) derived from Jupiter observations as a function of observing date. Error bars represent the detector-to-detector scatter quantified via the median absolute deviation. The horizontal line denotes the mean beam FWHM measured throughout the campaign, while the dashed lines indicate the corresponding standard deviation.
}
    \label{fig:gb06_beam_stability}
\end{figure}

Changes in the measured responsivity could in principle arise from variations in either the detector response or the effective beam. To test whether beam variations contribute significantly to the observed responsivity trends, Fig.~\ref{fig:gb06_beam_stability} presents the temporal evolution of the beam FWHM measured from the Jupiter observations.

The beam remains stable throughout the observing campaign, clustering at $30.1$\,arcmin with only minor scatter (0.5\,arcmin). No significant temporal trend is observed, indicating that beam variations are unlikely to be a major contributor to the observed responsivity variability. Table~\ref{tab:flatfield} in Appendix~\ref{app:relative_response} reports the single-KID FWHM across the month.

\section{RESPONSIVITY DEPENDENCE ON OPTICAL LOADING}
\label{sec:moon}

\subsection{Moon Observations}
\label{subsec:moon_vs_jupiter}

The Moon has played a central role in the characterization of GroundBIRD, serving as the primary reference for developing the pointing model.\cite{2024PTEP.2024b3F01S}
More generally, the Moon has traditionally been used as a bright microwave calibration source because of its high brightness temperature and frequent visibility (see, e.g., Refs.~\citenum{2024PASP..136k4505M,2024ApJS..273...26D,2026A&A...708A.173F}).

A detailed analysis of Moon-based responsivity measurements for GroundBIRD is presented in Ref.~\citenum{tomonaga:2025}. That study showed that the responsivity varies substantially with observing conditions and cannot be described by a single empirical relation to atmospheric PWV. In this work, we revisit those measurements to investigate the dependence of the detector response on optical loading and to compare the resulting responsivities with those derived from Jupiter observations.

\begin{figure}[ht]
    \centering
    \includegraphics[width=0.70\linewidth]{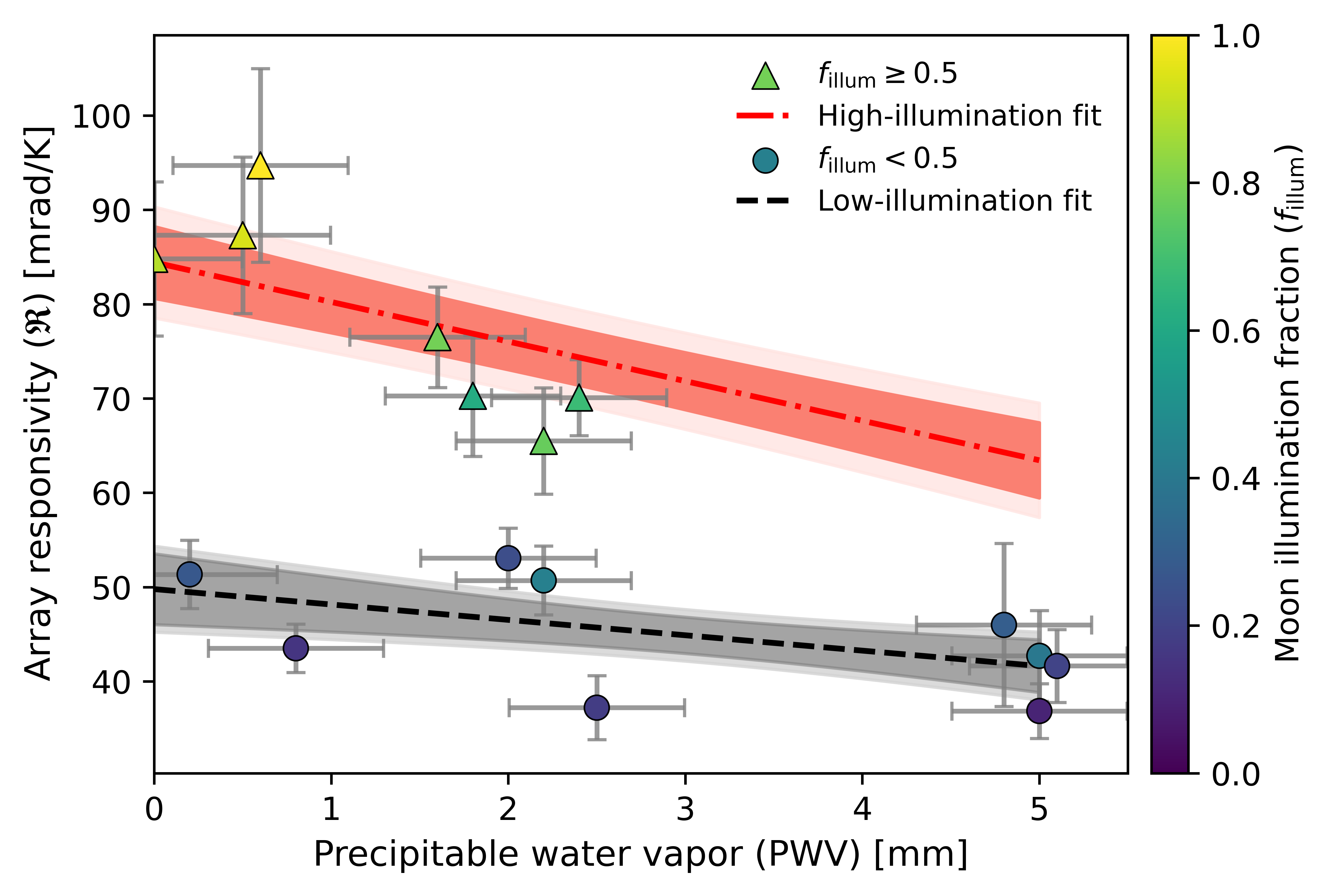}
    \caption{
Same as Fig.~\ref{fig:gb06_resp_pwv}, but for Moon observations. The marker color indicates the Moon illumination fraction, $f_{\rm illum}$, while circles (dashed fit) correspond to observations with $f_{\rm illum}<0.5$ and triangles (dash-dotted fit) to observations with $f_{\rm illum}\geq0.5$. Shaded bands denote the statistical fit uncertainty and the additional uncertainty associated with the PWV systematic error. Two distinct responsivity branches are observed, indicating that a single PWV-dependent calibration relation cannot describe the Moon responsivity.
}
    \label{fig:moon_resp_split}
\end{figure}

Unlike Jupiter, the Moon contributes a substantial fraction of the total optical load seen by the KIDs. As a result, Moon observations probe a different region of the detector operating condition and provide an opportunity to investigate the impact of total optical loading on the KID responsivity.
To investigate the origin of this behavior, we examine how the responsivity depends on the lunar illumination state, quantified by the illumination fraction ($f_{\rm illum}$), with $f_{\rm illum}=0$ at New Moon and $f_{\rm illum}=1$ at Full Moon. The observations are divided into two illumination regimes ($f_{\rm illum}<0.5$ and $f_{\rm illum}\geq0.5$), and the responsivity--PWV relation is fitted independently for each subset.
We use the responsivities computed in Ref.~\citenum{tomonaga:2025} and correct them for the corresponding atmospheric opacity.
The quoted responsivities are calculated taking into account the phase dependence of the Moon's temperature, using the model described in Ref.~\citenum{T_model}.
The results are shown in Fig.~\ref{fig:moon_resp_split}.

The high-illumination subset exhibits systematically higher responsivities than the low-illumination subset. The corresponding linear fits yield $\mathfrak{R}_0=84.4\pm3.5\,\mathrm{mrad/K}$ and $\alpha=-4.2\pm0.9\,\mathrm{mrad/K/mm}$ for $f_{\rm illum}\geq0.5$, compared with $\mathfrak{R}_0=50.0\pm3.5\,\mathrm{mrad/K}$ and $\alpha=-1.6\pm1.0\,\mathrm{mrad/K/mm}$ for $f_{\rm illum}<0.5$. 
The presence of two distinct responsivity branches confirms that a single PWV-dependent calibration relation cannot describe the Moon response.
Finally, the Moon-derived responsivities remain systematically higher than those obtained from Jupiter at comparable PWV (see Fig.~\ref{fig:gb06_resp_pwv}).
The discrepancy is particularly pronounced for the high-illumination subset, whose fitted responsivities exceed the Jupiter relation by more than $10\sigma$, whereas the low-illumination subset differs at $\sim2.5\sigma$ level. This behavior further supports the existence of illumination-dependent detector operating regimes.
Since the Moon contributes a substantially larger optical load than Jupiter, this offset cannot be explained by atmospheric loading alone and instead points to the existence of distinct detector operating regimes, as discussed in Sect.~\ref{sec:total_optical_load}.

\subsection{Variable-Component Antenna Temperature Dependence}
\label{sec:total_optical_load}

Under comparable atmospheric conditions, responsivities from the Moon and Jupiter show inconsistency.
This behavior suggests that when the source imposes high optical loading, we need to account for this distinct regime.
To explore this possibility, we analyze observations of the Moon and Jupiter using the variable-component antenna temperature ($T_{\rm A,tot}$) as an indicator of optical loading. This metric captures the portion of the optical load that varies with the observation, including contributions from the atmosphere ($T_{\rm A,atm}$) and the astronomical source ($T_{\rm A,source}^{\rm obs}$), while excluding the approximately constant contributions from the telescope and instrument optics.
A smaller second-order contribution may also arise from variations in the focal-plane temperature, which can slightly modify the KID operating point and contribute to the observed scatter.\cite{11429154}
The variable-component antenna temperature is, thus, defined as
\begin{equation}
T_{\rm A,tot} = T_{\rm A,atm} + T_{\rm A,source}^{\rm obs} \, ,
\end{equation}
where $T_{\rm A,source}^{\rm obs}$ represents the observed Moon or Jupiter contribution after atmospheric attenuation and beam coupling. 
For Jupiter, the beam dilution is computed assuming a point source, whereas for the Moon, the Gaussian beam is integrated over a uniform disk with the observed lunar angular diameter.
For the Moon antenna temperature, we employ the model described in Ref.~\citenum{2014MNRAS.439.2271H} at 145\,GHz.
In this work, we adopt a spatially uniform lunar brightness temperature across the entire disk, rather than implementing a spatially resolved brightness-temperature distribution. While latitude-dependent lunar microwave brightness-temperature maps have been derived at lower frequencies (3.0, 7.8, 19.4, and 37\,GHz; see Ref.~\citenum{2012Icar..219..194Z}), we are not aware of corresponding observational constraints at higher frequencies. Given that the GroundBIRD beam FWHM ($\sim30$\,arcmin) is comparable to the lunar angular diameter (29.6--32.1\,arcmin over the observations considered here), the resulting measurements effectively represent an average over a substantial fraction of the lunar surface. We therefore expect that any spatial variations in brightness temperature across the disk would contribute only a second-order correction, which lies beyond the scope of the present analysis.

\begin{figure}[h!]
\centering
\includegraphics[width=0.58\textwidth]{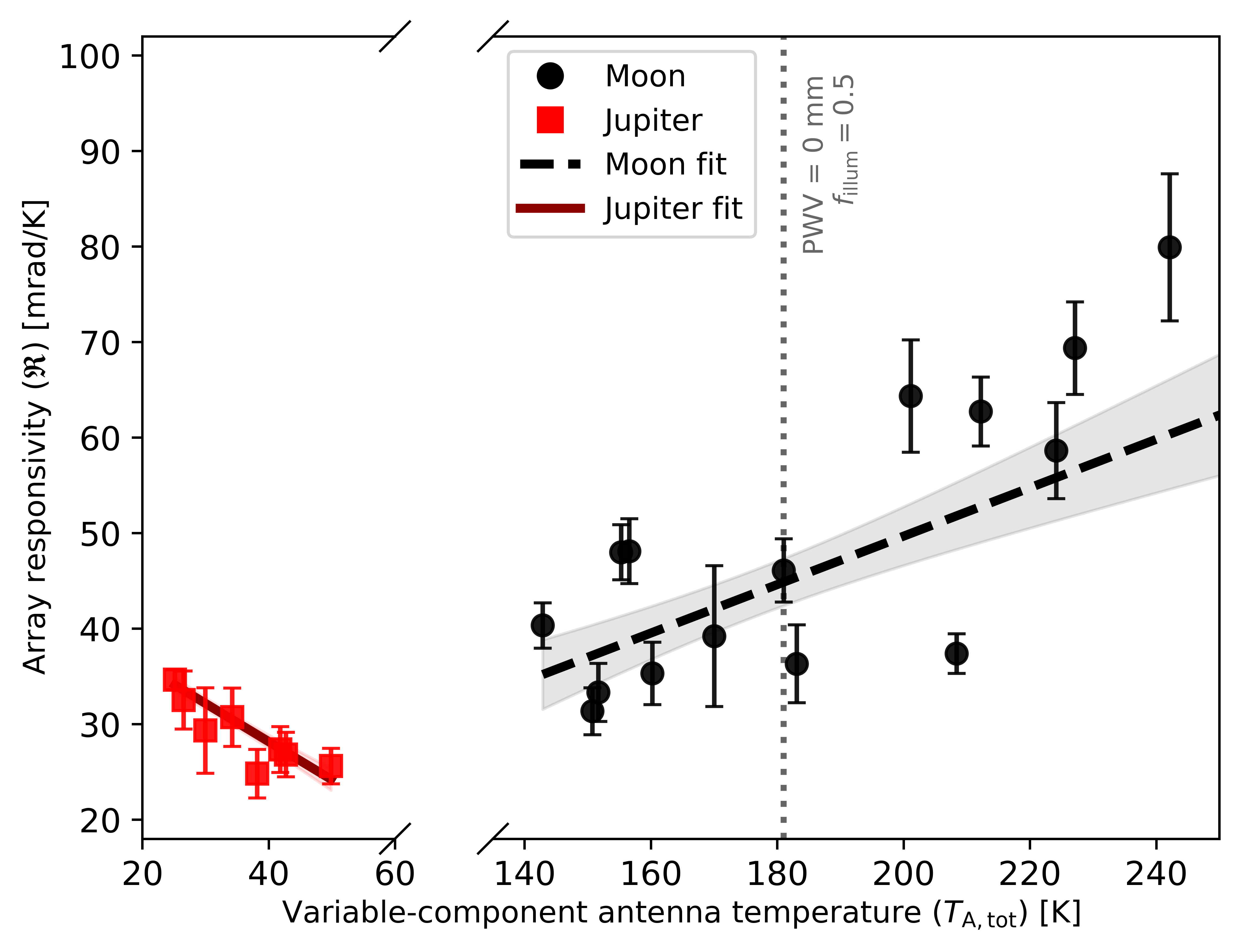}
\caption{
Array responsivity as a function of the variable-component antenna temperature for Jupiter and Moon observations. 
The solid lines represent linear regressions to the Jupiter and Moon measurements, while the shaded bands denote the corresponding $1\sigma$ uncertainties on the fitted relations.
The vertical dotted line marks $T_{\rm A,tot}=181$\,K, corresponding to PWV$=0$\,mm and an illumination fraction $f_{\rm illum}=0.5$.
}
\label{fig:moon_jupiter_total_loading}
\end{figure}

Assuming an average GB06 observing elevation of 65\,deg, the atmospheric contribution was quantified using a simplified single-layer atmospheric model. The zenith opacity was parameterized as a linear function of PWV, using coefficients obtained from the \texttt{am} atmospheric transmission code convolved with the GroundBIRD instrumental bandpass. The corresponding atmospheric emission was then calculated by adopting an effective atmospheric temperature of 283\,K, which is representative of the mean conditions above Teide Observatory \cite{5e31703e2999523690ffef95}.
The resulting variable-component antenna temperature is subsequently employed to compare the Moon and Jupiter responsivities on a common scale. Figure~\ref{fig:moon_jupiter_total_loading} presents the measured responsivity as a function of the variable-component antenna temperature for both Moon and Jupiter observations.

The inferred slopes are $-0.4\pm0.1\,\mathrm{mrad/K^{2}}$ for Jupiter and $0.3\pm0.1\,\mathrm{mrad/K^{2}}$ for the Moon, with corresponding intercepts of $44.2\pm2.1\,\mathrm{mrad/K}$ and $-1.0\pm14.3\,\mathrm{mrad/K}$, respectively. 
The opposite signs of the fitted slopes indicate that the two calibrators probe distinct KID operating regimes. This behavior is qualitatively reproduced by a simple resonator model, in which a sufficiently large optical load drives the detector substantially away from its nominal operating point, resulting in an increasingly off-resonance readout configuration.

Jupiter measurements are confined to low antenna temperatures and follow the PWV-dependent responsivity relation derived in Sect.~\ref{sec:responsivity_vs_pwv}, whereas Moon measurements extend to substantially higher antenna temperatures and exhibit systematically larger responsivities. This separation demonstrates that PWV alone does not provide a sufficient description of the detector response. Observations obtained under comparable atmospheric conditions yield markedly different responsivities when the total optical loading differs significantly. Alternative explanations, including beam-size variations (Sect.~\ref{sec:gb06_stability}) and focal-plane temperature fluctuations, cannot account for the amplitude of the observed offset. Instead, the data indicate that the detector responsivity is primarily governed by the operating point established by the total optical load. 
In conclusion, Jupiter observations probe the atmosphere-dominated operating regime relevant to routine observations, whereas Moon observations sample a distinct high-optical-loading, off-resonance operating regime.

\subsection{Jupiter as a Calibrator of the Atmosphere-Dominated Regime}
\label{sec:jupiter_as_calibrator}

Under representative observing conditions characterized by a PWV level of 4.3\,mm, the atmospheric contribution to the antenna temperature is $T_{\rm A,atm}\simeq36.0\,\mathrm{K}$ at the GroundBIRD observing elevation in the 145\,GHz band. In comparison, the observed contribution from Jupiter in this study is only $T_{\rm A,Jupiter}^{\rm obs}\simeq50\,\mathrm{mK}$. Jupiter therefore contributes less than 0.5\% of the total variable optical load, demonstrating that the atmospheric background dominates the detector operating point during Jupiter observations and remains representative of nominal survey conditions. 
In contrast, the observed contribution from the Moon exceeds the atmospheric loading by a factor of $>5$ during the observation, driving the detectors into a substantially different operating regime.\footnote{For the Moon observations considered in this work, the median atmospheric contribution is $T_{\rm A,atm}\simeq29.5\,\mathrm{K}$ and the median observed source contribution is $T_{\rm A,Moon}^{\rm obs}\simeq163.8\,\mathrm{K}$, yielding $T_{\rm A,Moon}^{\rm obs}/T_{\rm A,atm}\simeq5.6$.}

For dedicated observations of bright calibration sources, including Jupiter and the Moon, scans are typically extended to one hour, whereas nominal survey observations employ 10-minute tuning cycles.
This reflects a trade-off between observing efficiency and calibration stability: longer scans increase the available on-source integration time, but reduce the ability to track changes in atmospheric loading and detector operating conditions through frequent retuning (the tuning process takes $\sim40$\,s). Over a one-hour interval, the measured PWV fluctuations at the GroundBIRD site are typically $\sim0.7$\,mm, corresponding to a variation of $\sim3$\,K in atmospheric loading. The longer interval between retuning operations therefore increases the impact of atmospheric fluctuations on the detector operating point and, consequently, on the responsivity calibration. By extrapolating the measured one-hour PWV stability to the nominal 10-minute tuning cycle, we estimate an atmospheric-loading variation of $\sim1.2$\,K.
Under nominal observing conditions, the KID operating point is therefore dominated by atmospheric-loading fluctuations, while the observed Jupiter signal contributes only $\sim50$\,mK, about a factor of $\sim$24 less than the expected atmospheric-loading variation between successive retuning operations.

Jupiter thus constitutes a suitable calibrator for measuring the on-sky responsivity under representative observing conditions, without significantly perturbing the nominal KID operating point. Since the optical load continues to be dominated by atmospheric emission, the ultimate precision of the calibration is fundamentally limited by fluctuations in the atmospheric background.

\section{CONCLUSIONS}
\label{sec:conclusions}

This work presents the first on-sky responsivity calibration model for the GroundBIRD experiment. 
Using Jupiter observations acquired with twelve KIDs selected for their consistently repeatable detections throughout the observing campaign, we derive the first empirical relation between detector responsivity and atmospheric PWV for this instrument.

The selected KID ensemble exhibits stability on month-long timescales at the level of better than 20\%, while the responsivity decreases by approximately 30\% across the sampled PWV range of 1.8--4.4\,mm. These results demonstrate that the responsivity is not constant but instead varies with atmospheric conditions through changes in the detector operating point.
The responsivity trend measured with Jupiter is consistent with the expected behavior of KIDs that record the phase signal under increasing optical loading, and it can be described by a PWV-dependent linear relation up to PWV$\simeq4.4$\,mm. Within this regime, the calibration constrains the array responsivity with a precision better than 20\% under typical observing conditions.

A comparison with Moon observations further reveals the existence of two distinct detector operating regimes: an atmosphere-dominated regime probed by Jupiter observations and a high-optical-loading regime probed by the Moon. The opposite responsivity trends observed indicate that the detector response cannot be parameterized solely in terms of atmospheric conditions.
Instead, the responsivity is governed by the detector operating point, which evolves with the total optical loading as the resonator is driven progressively off resonance.

The present analysis demonstrates that Jupiter is a suitable primary calibrator for establishing a robust on-sky responsivity calibration for GroundBIRD. 
More broadly, the results highlight the importance of accounting for optical-loading effects when calibrating KID-based instruments.
The calibration framework developed in this work provides a foundation for extending the analysis to the full GroundBIRD focal plane and for further validation using additional astronomical targets and independent measurements of atmospheric loading.

\appendix    

\section{Per-Detector Calibration Parameters}
\label{app:relative_response}

Table~\ref{tab:flatfield} reports the relative response, median responsivities, and beam FWHM values derived from Jupiter observations for the twelve selected KIDs (see Sects.~\ref{sec:relative_response} and \ref{sec:gb06_stability}).

\begin{table}[ht]
\centering
\caption{
Relative response ($F_{\rm k}$), median responsivity $\langle\mathfrak{R}_{\rm k}\rangle$, and median beam full width at half maximum (FWHM) derived from Jupiter observations for the selected KIDs. The quoted uncertainties correspond to the temporal dispersion estimated from the scaled median absolute deviation of the individual measurements throughout the observing campaign.
}
\label{tab:flatfield}
\begin{tabular}{ccccc}
\hline
KID ID & $F_{\rm k}$ & $\langle \mathfrak{R}_{\rm k} \rangle$ [mrad/K] & FWHM [arcmin] \\
\hline
0  & $1.18\pm0.08$ & $33.5\pm6.1$ & $29.9\pm1.4$ \\
1  & $0.89\pm0.05$ & $25.4\pm2.2$ & $29.9\pm2.6$ \\
5  & $0.73\pm0.07$ & $20.9\pm4.4$ & $31.5\pm1.6$ \\
6  & $1.18\pm0.06$ & $33.0\pm1.9$ & $29.7\pm0.9$ \\
7  & $1.12\pm0.10$ & $31.2\pm3.6$ & $30.2\pm1.2$ \\
8  & $1.12\pm0.03$ & $33.3\pm6.3$ & $30.3\pm1.2$ \\
9  & $1.15\pm0.14$ & $33.9\pm6.8$ & $29.7\pm1.9$ \\
10 & $1.20\pm0.08$ & $35.4\pm6.7$ & $30.7\pm1.7$ \\
11 & $0.46\pm0.07$ & $13.1\pm0.6$ & $29.6\pm2.3$ \\
14 & $0.70\pm0.08$ & $20.2\pm3.9$ & $30.0\pm1.6$ \\
15 & $0.76\pm0.12$ & $21.7\pm4.6$ & $31.1\pm2.7$ \\
16 & $0.42\pm0.07$ & $13.2\pm2.9$ & $28.5\pm2.3$ \\
\hline
\end{tabular}
\end{table}

The measured responsivity spread is consistent with the expected detector-to-detector variations arising from differences in resonator characteristics, optical coupling, and readout response. Despite these variations in absolute responsivity, the relative responses remain stable throughout the observing campaign, with typical temporal variations below 10\% and no detector exceeding 17\%.
The beam FWHM remains stable throughout the campaign, with median values ranging from 28.5-- 31.5\,arcmin. The mean per-detector fractional scatter is approximately 6\%, and no detector exhibits FWHM variations exceeding 9\%.

\acknowledgments 
 
This work was supported by the GroundBIRD collaboration.
This work was supported by MEXT KAKENHI Grant Number JP18H05539 and JSPS KAKENHI Grant Numbers JP15H05743, JP20K20927, JP20KK0065, JP21H04485, JP21K03585, JP22H04913, and JP24H00224, JSPS Bilateral Program Numbers JPJSBP120219943 and JPJSBP120239919, and also supported by JSPS Core-to-Core Program JPJSCCA20200003.
This article was partially supported by the Korea University Research Grant and the Research Support Grant RS-2022-NR068913. We are partially supported by high-speed KREONET provided by KISTI.
Partial financial support was provided by the Spanish Ministry of Science and Innovation MCIN/AEI/10.13039/501100011033 under the project PID2023-151567NB-I00.
The on-site PWV measurements were taken with a radiometer manufactured by Furuno Electric Co., Ltd.
We thank Victor González Escalera, \'{A}ngeles P\'{e}rez de Taoro (the Instituto de Astrof\'{i}sica de Canarias), and the staff of Teide Observatory for supporting the maintenance and operation of GroundBIRD, MOT, MANTIN, and TOT.
AF thanks Shunichi Nichi for insight into the Furuno monitor.

\bibliography{report} 
\bibliographystyle{spiebib} 

\end{document}